%% file: main.tex
\documentclass{article}

\usepackage[preprint]{neurips_2019}

\usepackage[utf8]{inputenc} 
\usepackage[T1]{fontenc}    
\usepackage[colorlinks=true,urlcolor=blue,citecolor=magenta]{hyperref}       
\usepackage{url}            
\usepackage{booktabs}       
\usepackage{amsfonts}       
\usepackage{nicefrac}       
\usepackage{microtype}      
\usepackage{mathtools}
\usepackage{algpseudocode}

\author{%
   Paul Bertin \\
   Mila, Universit\'{e} de Montr\'{e}al \\ Montr\'{e}al, Canada
   \And
   Mohammad Hashir \\
   Mila, Universit\'{e} de Montr\'{e}al \\ Montr\'{e}al, Canada
    \And
    Martin Weiss \\
    Mila, Universit\'{e} de Montr\'{e}al \\ Montr\'{e}al, Canada
    \And 
    Vincent Frappier \\
    Mila, Universit\'{e} de Montr\'{e}al \\ Montr\'{e}al, Canada
    \And 
    Theodore J. Perkins \\
    Ottawa Hospital Research Institute \\ University of Ottawa \\ Ottawa, Canada
    \And 
    Genevi\`{e}ve Boucher \\ 
    Institute for Research in Immunology and Cancer \\ Universit\'{e} de Montr\'{e}al \\ Montr\'{e}al, Canada
    \And 
    Joseph Paul Cohen \\ 
    Mila, Universit\'{e} de Montr\'{e}al \\ Montr\'{e}al, Canada
}

\usepackage{longtable}
\usepackage{float}
\usepackage{wrapfig}

\usepackage{booktabs}
\usepackage{multirow}
\usepackage{afterpage}

\usepackage{microtype}
\usepackage{graphicx}
\usepackage{subfig}
\usepackage{amsmath}

\title{Analysis of Gene Interaction Graphs as Prior Knowledge for Machine Learning Models}

\begin{document}

\maketitle

\begin{abstract}
Gene interaction graphs aim to capture various relationships between genes and can represent decades of biology research. When trying to make predictions from genomic data, those graphs could be used to overcome the curse of dimensionality by making machine learning models sparser and more consistent with biological common knowledge. 
In this work, we focus on assessing how well those graphs capture dependencies seen in gene expression data to evaluate the adequacy of the prior knowledge provided by those graphs. We propose 
a condition graphs should satisfy to provide good prior knowledge and  test it using `Single Gene Inference' tasks. 
We also compare with randomly generated graphs, aiming to measure the true benefit of using biologically relevant graphs in this context, and validate our findings with five clinical tasks. We find some graphs capture relevant dependencies for most genes while being very sparse. Our analysis with random graphs finds that dependencies can be captured almost as well at random which suggests that, in terms of gene expression levels, the relevant information about the state of the cell is spread across many genes. Our source code is available on \href{https://github.com/Bertinus/gene-graph-analysis}{GitHub} 

\end{abstract}

{\let\thefootnote\relax\footnote{Corresponding authors: \texttt{bertinpa@mila.quebec}, \texttt{mohammad.hashir.khan@umontreal.ca} or \texttt{joseph@josephpcohen.com}}}

\setcounter{footnote}{0}
\input{sections/introduction.tex}
\input{sections/methods.tex}

\input{sections/results.tex}

\section*{Funding}
This work is partially funded by a grant from the Fonds de Recherche en Sante du Quebec and the Institut de valorisation des donnees (IVADO).  This work utilized the supercomputing facilities managed by Mila, NSERC, Compute Canada, and Calcul Quebec. We also thank NVIDIA for donating a DGX-1 computer used in this work.

\section*{Acknowledgements}
We thank Francis Dutil for sharing code. We also thank Yoshua Bengio and Mandana Samiei for their useful feedback. 

\bibliographystyle{natbib}
\bibliography{bibliography}


\input{sections/appendix.tex}

\end{document}

%% file: sections/introduction.tex
\section{Introduction}

A major challenge in using machine learning on gene expression data is overcoming the curse of dimensionality \citep{yeung2001principal,kim2003subsystem,lee2008investigating,zuber2009gene,heimberg2016low}. The number of samples in most datasets is typically much smaller than the number of genes. Making predictions based on gene expression can be problematic as the model would tend to overfit and learn the noise and spurious correlations in place of any biologically relevant patterns. 

Penalizing model complexity can reduce this problem \citep{gustafsson2005constructing,cawley2006gene,ma2009regularized,simon2013sparse,min2018network} and can simultaneously perform feature selection, identifying relevant genes. 
However, gene selection is not generally stable unless special algorithms are used \citep{he2010stable}. Moreover, regularization typically emphasizes some general notion of sparsity rather than any specific prior biological knowledge. Some recent works have looked at incorporating biological knowledge into regularization, e.g. based on pathways relationships \citep{simon2013sparse,min2018network}.

In this work, we focus on gene interaction graphs as a form of prior biological knowledge for machine learning models. Many groups have developed a number of gene-interaction graphs, structuring domain knowledge from different areas of molecular biology \citep{Ogris2018,Warde-Farley2010,Himmelstein2015,Himmelstein2017-vq,Lee2011, Hwang2019-gb,Subramanian2017_s,liu2015regnetwork,Kanehisa2017,Szklarczyk2019}. These graphs can represent any number of different biological, molecular, or phenomenological relationships such as protein-protein interactions, transcriptional regulation, transcriptional co-regulation, co-expression at the mRNA or protein levels, etc.
Gene interaction graphs can be used with machine learning algorithms as a proxy for biological intuition to leverage decades of biology research \citep{Zhang2017}. They can act as a biological prior on machine learning techniques to automate feature importance \& selection and help to overcome the curse of dimensionality.
For example, network-based linear regression \citep{Li2008, Min2016} regularizes the weights of a linear model based on the connectivity of the nodes found in an interaction graph. Preliminary work by \citet{Rhee2018} and \citet{Dutil2018} found that the same can be done for non-linear models. This would be useful in the majority of tasks where gene expression or single-nucleotide polymorphism data is the input. As these graphs were not developed as an input for machine learning applications, there is value in investigating whether they can aid machine learning algorithms. Thus there is need for a systematic approach to assess the value of this prior biological knowledge for machine learning pipelines.

We propose a method to quantitatively evaluate the prior knowledge provided by gene interaction graphs: we formulate a condition that a graph should satisfy to provide good prior knowledge for machine learning algorithms, and then test to which extent this condition holds. Seven major gene interaction graphs created by different research groups (which we refer to as `curated graphs') were tested. We compare curated graphs with randomly generated graphs, aiming to measure the true benefit of using biologically relevant graphs in this context. 
We also perform experiments on clinical prediction tasks: we compare the performance of models provided with different graph-based prior knowledge in order to validate our evaluation method.

With this work, we aim to gain a deeper insight into the behavior of machine learning pipelines that make use of graph-based prior knowledge when applied to gene expression data. This effort can help in improving interpretability which is very important for genomics as it is a domain where we have relatively limited intuition compared to images or text. Having more interpretable models could provide a ``research gradient'' to biologists allowing them to focus on specific subgroups of genes, which could lead to a fruitful feedback loop between biological experiments and machine learning predictions. Interpreting those models could also help in generating new hypotheses that may be validated with experiments. Our approach is different from those of most module detection methods \citep{Saelens2018} which infer modules of genes that work together from gene expression data. They typically aim at better understanding biological mechanisms. On the other hand, our method was developed to facilitate the use of pre-established biological knowledge to exploit more efficiently gene expression data and make accurate predictions.

%% file: sections/methods.tex
\section{Materials and methods}

\subsection{What is a ``good'' graph-based prior knowledge?}

For a set of gene expressions $\{g_i\}_{i \in [1...N]}$ where $N$ is the number of genes, the joint probability of the expressions is denoted by $P( g_1, ..., g_N)$. We hypothesize that there is a true causal (directed) graph $G$ that generated this distribution: each gene expression was generated by a function $f$ such that $g_i \doteq f(\mathit{PA}_i, \epsilon)$ where $\mathit{PA}_i$ refers to the set of expressions of the parents of node $i$ and $\epsilon$ is some random noise.

\begin{figure}[!h]
    \centering
    \includegraphics[width=0.4\columnwidth]{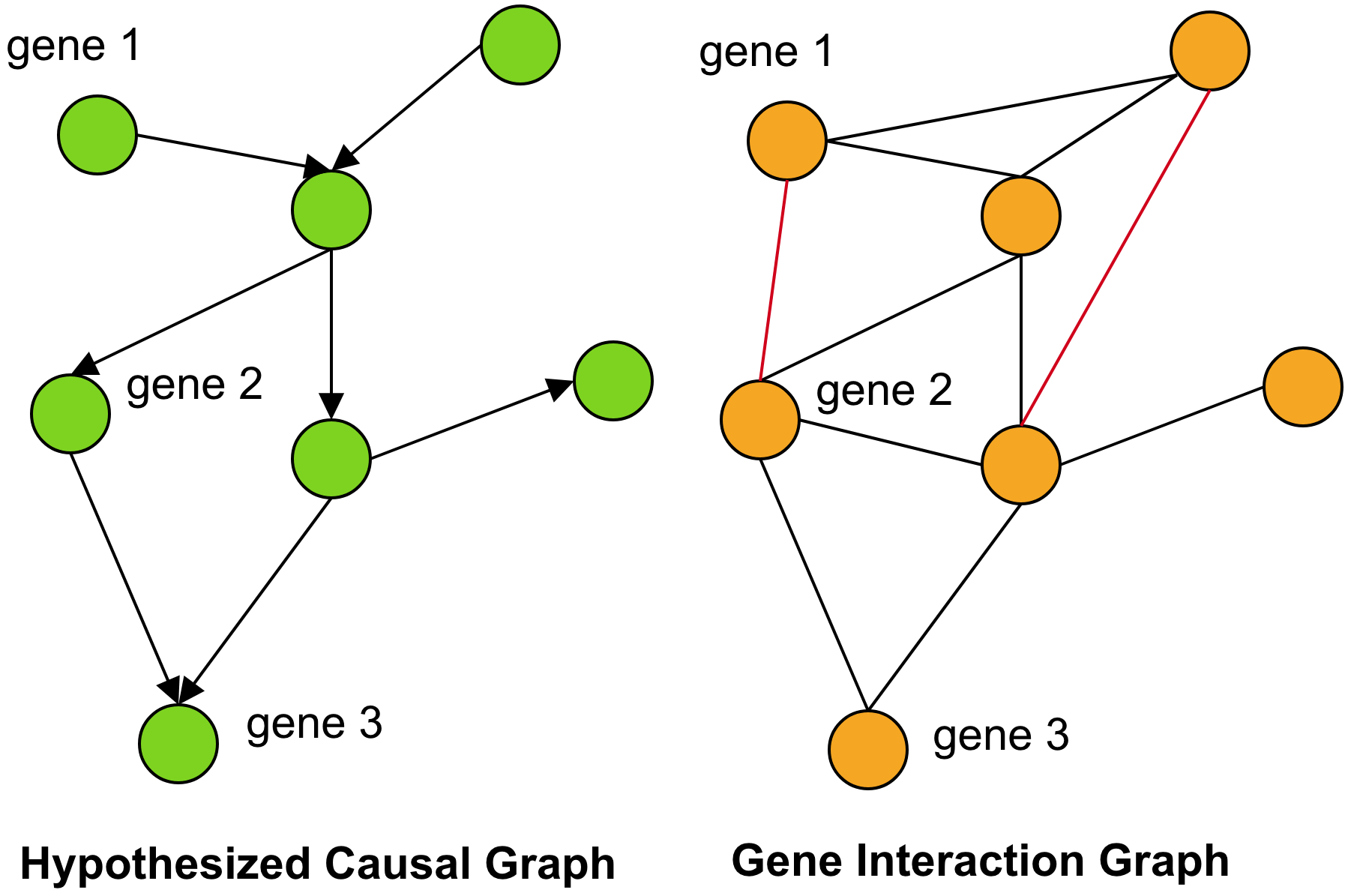}
    \caption{We define a ``good'' prior knowledge as an interaction graph $G$ that covers the moral (undirected) graph equivalent to the hypothesized causal graph, \textit{i.e.} edges in $G$ cover all edges found in the hypothesized \textit{true} causal graph, and parents with a common child are connected in $G$. $G$ could also contain other \textit{spurious} edges (in red) while still satisfying the inclusivity property, as in this example.}
    \label{fig:diagram_graphs}
\end{figure}

In order to provide ``good'' prior knowledge, we would like our gene interaction graph to be equivalent to the hypothesized \textit{true} causal graph. Figure \ref{fig:diagram_graphs} illustrates the relationship between the two graphs.

\textbf{Definition} \hspace{3pt} A ``good'' prior knowledge is a gene interaction (undirected) graph $G$ which covers the moralized counterpart of the \textit{true} causal graph $H$, \textit{i.e.} if an edge $i \rightarrow j$ exists in $H$, the edge $\{i,j\}$ should exist in $G$, and if two nodes have a common child in $H$, an edge should exist between them in $G$.

\textbf{Inclusivity Property} \hspace{3pt} If an interaction graph $G$ is a ``good'' prior knowledge, then for any gene node $i$, the Markov blanket of $i$ in the \textit{true} causal graph is contained in the set of neighbours of node $i$ in $G$. Equivalently, if $G$ is a ``good'' prior knowledge, the following holds for all gene nodes $i$: 
\begin{equation}
    P(g_i | \overline{g_i}) = P(g_i | neighbours_i) 
    \label{eq:1}
\end{equation}
where $\overline{g_i} = \{g_j\}_{j \ne i}$ is the set that contains every gene expression except the $i^{th}$ one, and $neighbours_i = \{g_j | \exists \text{ edge } \{i, j\} \text{ in graph } G \}$

If the equality (Eq. \ref{eq:1}) holds for a gene $i$, it means that the conditional probability of $g_i$ given all the other genes only depends on the first degree neighbours of $g_i$. 
Note that the inclusivity property does not ensure that the interaction graph has no spurious edges. An edge $\{i, j\}$ in the interaction graph is called spurious if it does not exist in the moralized counterpart of the hypothesized \textit{true} causal graph. Also, the detection of spurious edges is not our main concern as we deal with fairly sparse graphs. The goal of this work is to identify graphs that are sparse while still satisfying the inclusivity property.

\subsection{Method}\label{sec:methods}

As there is no direct way to test the equality (Eq. \ref{eq:1}), we model the conditional probability of the expression of gene $g_i$ given all the other genes with a neural network. This task of predicting a gene expression value given a set of other gene expressions is similar to the Single Gene Inference task formulated in \citet{Dutil2018}, which was inspired by \cite{Chen2016} and \cite{Subramanian2017_s}. 

\begin{figure}[!h]
    \centering
    \includegraphics[width=0.4\columnwidth]{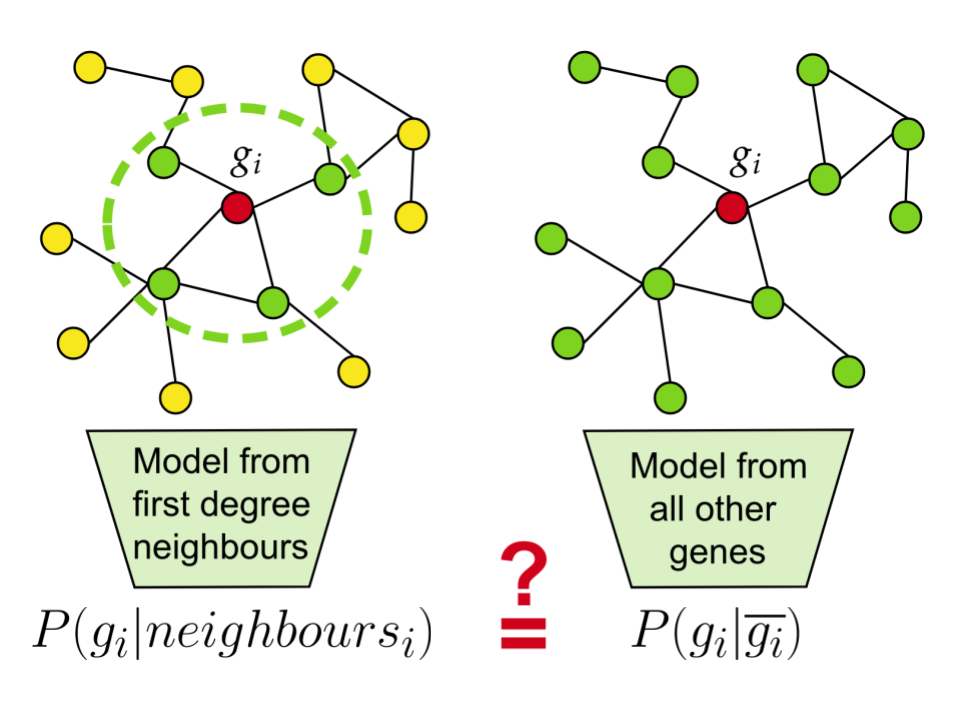}
    \caption{Given a graph and a target gene $g_i$, we train two different models taking first degree neighbours and all genes as input, respectively. They model $P(g_i | neighbours_i)$ and $P(g_i | \overline{g_i})$ respectively. We can compare the performance of the two models to infer the quality of the prior knowledge provided by the graph for $g_i$}
    \label{fig:pipeline}
\end{figure}

For each gene \textit{i}, we train two different models that try to predict its expression level $g_i$. The first model takes all the other genes as input and the second takes only the first degree neighbours as input. 
If the equality holds for $g_i$, both models should achieve similar performance and approximate the conditional probability equally well. We can even expect slightly better performance in the second model as signal is lower dimensional and supposed to contain only relevant information. Conversely, if the equality does not hold for $g_i$, then we expect the second model to achieve poorer performance as it will be provided with incomplete information. Figure \ref{fig:pipeline} shows a schematic view of our method.

We restrict the prediction of $g_i$ to a binary classification task to simplify interpretation of the results. The alternative is to define a regression task, but depending on the pattern of expression of the gene, its range, its level of noise or any other specificity, the regression metric of a given gene (\textit{e.g.} R-squared) can be arbitrarily high even for a \textit{good} fit. A reduction of performance in some genes can be missed when looking at global statistics of the regression task.
The key point here is that we are interested in aggregated results (\textit{e.g.} mean over all genes) which requires metrics that are comparable between genes. We believe that the area under the ROC curve (AUC) of a binary classification task matches those requirements. Another reason we use classification is that we are not aiming to precisely model gene expression patterns; rather we intend to compare the two models to know whether the equality (Eq. \ref{eq:1}) holds or not.

\subsection{Datasets}
We perform our analysis with gene expression data not only  from \textit{healthy} cells, but also from \textit{cancerous} cells where biological processes might somehow be perturbed, giving us an idea of the usefulness of using gene interaction graphs in different contexts. We use the TCGA dataset\footnote{version 2016-08-16} \citep{CancerGenomeAtlasResearchNetwork2013} where values are $log_2(x+1)$ transformed RSEM values \citep{li_rsem:_2011} and the GTEx dataset\footnote{\url{https://cbcl.ics.uci.edu/public_data/D-GEX/GTEx_RNASeq_RPKM_n2921x55993.gctx}} \citep{lonsdale2013genotype} where values are RPKMs \citep{mortazavi_mapping_2008}. Both those datasets are public and well-studied. The TCGA PANCAN database spans multiple tissues and measures 20,530 gene expressions for 10,459 samples; most samples come from cancer biopsies but many healthy examples are also included. The GTEx dataset consists of samples mostly from healthy subjects and has a higher amount of genomic features (34,218 features) but only for 2,921 samples. We normalize both datasets by the mean (gene-wise) for our analysis.

\subsection{Graphs}

We evaluate six existing graphs covering a variety of relationships in the genome, namely FunCoup, GeneMANIA, Hetionet, HumanNet, RegNet, and StringDB. In addition, we create a graph from the Landmark gene set \citep{Subramanian2017_s}. We refer to these seven graphs as `curated' graphs. We also create graphs with \textit{n} neighbours, selected randomly, for each gene in a given dataset which we call random R-\textit{n} graphs.

We made several simplifications to be able to use all the graphs in a similar manner. The naming convention of genes differed from a graph to another; we used the HUGO gene symbols as a common schema and generated conversion maps from the HUGO Gene Nomenclature Committee website\footnote{\url{https://www.genenames.org/}}. We renamed the nodes in each graph to their HUGO equivalent to ensure uniformity and have an easier comparison between graphs. Further, we considered edges to be binary (present or absent) and didn't take any weights that might have been associated with them.

\paragraph{\textbf{Fully Connected}}
We consider that the baseline model taking all gene expressions except the target as input could be represented as a `fully connected' graph. Indeed, in the fully connected graph, the first degree neighbours of a gene are all the other genes.

\paragraph{\textbf{FunCoup}}
FunCoup \citep{Ogris2018} used a Bayesian algorithm and orthology based information transfer in order to infer functionnal associations. It builds the interaction networks for 17 organisms; we use the network for humans. The edges in FunCoup have been given confidence scores for supporting different kinds of interactions such as metabolic, signaling, complex and physical protein interactions. It contains 5,505,787 edges covering 16,765 gene nodes.

\paragraph{\textbf{GeneMANIA}}
\citep{Warde-Farley2010} is an interaction network built from hundreds of datasets collected from GEO\footnote{\url{https://www.ncbi.nlm.nih.gov/geo/}}, BioGRID\footnote{\url{https://thebiogrid.org/}}, Pathway Commons\footnote{\url{https://www.pathwaycommons.org/}} and I2D\footnote{\url{http://ophid.utoronto.ca/ophidv2.204/}}. Ridge regression as well as a label propagation algorithm were used to predict gene functions. It contains 264,657 edges covering 16,297 gene nodes.

\paragraph{\textbf{Hetionet}}
\citep{Himmelstein2015, Himmelstein2017-vq} is a heterogeneous biological network consisting of multiple node and edge types which was created using public biological databases. It contains around 47,000 nodes of 11 types such as genes, compounds, pathways, etc., with more than 2 million edges. The gene nodes in Hetionet were curated from Entrez Gene and have three types of edges between them: interaction, co-variation and regulation.
In our analysis, we combine all these edges and create an undirected graph to represent the gene sub-network in Hetionet. This graph has 471,046 edges covering 17,681 gene nodes.

\paragraph{\textbf{HumanNet}}
\citep{Lee2011, Hwang2019-gb} contains functional associations that were inferred from many sources such as protein-protein interactions (PPIs), gene co-expression, protein co-occurrence, and genomic contexts. Further, the original graphical database was improved by augmenting its data sources such as the source texts for the co-citation network and interologs from other species. 
HumanNetV2 has an inclusive hierarchy in its constituent gene networks. We use the highest level of HumanNetV2 called HumanNet-XN in our analysis; it contains 476,399 edges covering 16,243 gene nodes.

\paragraph{\textbf{RegNetwork}}
\citep{liu2015regnetwork}, is composed of experimental and predicted up/down regulating effects between genes and includes information from KEGG \citep{Kanehisa2017}. RegNetwork contains 247,848 edges covering 7,220 gene nodes.

\paragraph{\textbf{STRINGdb}}
\citep{Szklarczyk2019} is a network of Protein-Protein Interactions. It includes several types of interactions among which are co citation in research papers (text mining), co-expression, gene neighbourhood, and involvement in common metabolic pathways.  For each protein, we retrieve the corresponding protein-coding gene(s), in order to create a gene-gene interaction network. StringDB contains different sub-graphs depending on the interaction type; we use the co-expression graph and the entire graph in our analysis. The entire graph contains 5,703,510 edges between 18,851 gene nodes and the co-expression graph contains 2,945,888 edges between 18,520 gene nodes.

\paragraph{\textbf{Landmark Graph}}
We generate a separate graph based on the Landmark genes \citep{Subramanian2017_s}. Those 978 genes were chosen to optimally recover the observed connections seen in the pilot Connectivity Map dataset. We build a graph in which each gene in a given dataset is connected to the set of the 978 landmark genes. The latter set of genes are themselves connected together, forming a clique. We refer to this graph as the \textit{Landmark graph}. 

\paragraph{\textbf{Random R-\textit{n} graphs}}
We generated random graphs with a fixed number \textit{n} of randomly sampled neighbours from the set of genes in the dataset. For each target gene in the dataset, we sample \textit{n} other genes from the dataset and connect them to the target gene node to create an \textbf{R-\textit{n} graph}. We created 15 such graphs with \textit{n} varying between 10 and 10,000.

\subsection{Metrics on graphs}

In this section, we introduce two graph metrics which would aid in the evaluation of a graph as prior knowledge for machine learning models.
Different graphs have different number of nodes and we are interested in graphs that contain a good proportion of the genes in a given dataset as nodes. As the final purpose is to use the graph to include prior knowledge in machine learning models, it is important to have a graph with sufficient coverage of all genes present in a dataset as opposed to a small subset of well-studied genes.
Moreover, distribution of degrees also vary significantly across multiple graphs.
 
We define two metrics called \textit{coverage} and \textit{sparsity} of a graph to have a clearer picture of how graphs can differ from each other. The coverage is the percent of the genes in the dataset that are present in the graph as nodes. The sparsity is a measure of the reduction in edges of a graph relative to the fully connected graph. It is calculated by subtracting the ratio of the number of edges in a graph to the number of edges in the fully-connected graph (which is dataset-dependent) from 1 and then represented as a percent. 
\begin{align*}\label{eq:coverage}
    \textit{Coverage} &= \frac{n(D\cap G)}{n(D)} \times 100\\
    \textit{Sparsity} &= \left( 1 - \frac{E_{g}}{n(D) \times n(D)} \right) \times 100
\end{align*}
where \textit{D} and \textit{G} are the sets of the genes present in a particular dataset and graph respectively, $E_{g}$ is the number of edges in a sub-graph of the graph that contains only the genes present in the dataset, i.e., the intersection set $D\cap G$, and $n(\cdot)$ represents the cardinality of the set, i.e., the total number of elements in the set.

\subsection{Experimental Setup}

In order to predict the over- or under-expression of the target gene, we began by binarizing its expression values based on that gene's mean expression. Then, two multi-layer perceptrons (MLPs) were trained to predict the target gene expression level using the two types of inputs mentioned in Section \ref{sec:methods}: all the genes (the baseline) and only the first degree neighbours. The AUC was computed on a test set after training. If the target gene had no neighbours in the graph, an AUC of 0.5 was assigned because the absence of any input features made the prediction a random guess.

We performed these experiments for all the seven curated graphs and 15 R-\textit{n} graphs with both datasets and all the genes in each dataset. We ran three trials for each combination of a gene, graph and dataset and averaged all metrics across the trials for a robust evaluation. To ensure small overlapping of example sets between trials, we used 3000 samples for TCGA and 1500 samples for GTEx with equal splits between the training, testing and validation sets. The data and also the gene neighbours for the R-\textit{n} graphs were randomly sampled for every trial but the data remained the same for every gene regardless of graph.

\paragraph{\textbf{Protocol}}
For all datasets, graphs and genes, we:
\begin{enumerate}
\item Binarize the target gene expression values with the threshold based off the mean expression for the target gene
\item Find neighbours of the gene in the graph. If the gene is not present in the graph or has no neighbours, an AUC of 0.5 is assigned and the next steps are skipped
\item Fit a prediction model on a subset of the dataset corresponding to the first-degree neighbourhood ($\times$3 trials)
\item Compute and record AUC of the predictions ($\times$3 trials)
\end{enumerate}

\subsection{Model configuration}

We utilized a mutlilayer perceptron (MLP) with a single hidden layer of 16 neurons and ReLU activation functions. The binary cross-entropy loss was used with an Adam optimizer and a learning rate of 0.001 for the FunCoup, Hetionet, and fully-connected graphs and $7\mathrm{e}^{-4}$ for the rest, on all datasets. The weight decay parameter was set to $1\mathrm{e}^{-8}$. These hyperparameters were obtained with a search over the different graphs and datasets and 20 genes. We used early stopping based off the AUC computed on a validation set.

The MLP achieved slightly better performance than logistic regression ($+5\mathrm{e}^{-3}$ AUC on average over 20 genes). $L_1$ regularization was not used as it did not improve performance.

%% file: sections/results.tex
\section{Results}\label{evalrawauc}

We present results showing the performance of our models, as well as aggregated statistics computed for the different curated graphs. We then compare with random graphs and evaluate the curated graphs on clinical tasks.

For our analysis, we defined three distinct types of sets of gene for each dataset. The first was \textbf{All Genes} which consisted of all the genes in the dataset and was the same for all graphs. The second was \textbf{Covered Genes} which referred to the set of genes in the dataset covered by a given graph. Each graph's set of Covered Genes was unique. We report several results on both these sets.
We decided to do this because our protocol of assigning an AUC of 0.5 for uncovered genes makes the performance of graphs artificially poorer when they have low coverage. It would not be equitable to compare different graphs only on their covered genes as they cover different sets of genes, which could include genes that are easier or harder to predict. We thus choose to report results on both sets.
We also defined a third type of set of genes called \textbf{Intersection Genes}. This consisted of the set of genes in the dataset that were common to all graphs except RegNet. We chose not to include RegNet as we wanted this set to have a high cardinality and RegNet covers a relatively small number of genes (7,220). The Intersection Genes set had 14,445 and 14,270 genes for TCGA and GTEx respectively.

\begin{table*}[!h]
\centering
\caption{Statistics for each graph and dataset. \textbf{Coverage} is the percentage of genes in the dataset that are represented as nodes in the graph. \textbf{Sparsity} represents the percentage of missing edges in the graph compared to the fully connected graph. The \textbf{Covered Genes AUC}  and the \textbf{All Genes AUC} refer to the average AUC achieved by the graph, respectively, on only its covered genes and on the entire dataset (after adding uncovered genes with an AUC of 0.5). The \textbf{Improvement} is computed with respect to the fully-connected baseline for both the covered genes and all genes.
\textbf{Per-gene AUC STD} refers to the per-gene standard deviation of AUCs across the three trials (averaged over all genes for a given graph and dataset). All uncertainties are computed as the standard deviation across the three trials.}
\label{tab:coverage}
\resizebox{\textwidth}{!}{ 
\begin{tabular}{llrrrrrrrr}
	\toprule
	                                               &                               &   \parbox{30pt}{\footnotesize\centering \hspace{6pt} Fully\\Connected} & GeneMANIA & RegNet & HNetV2 &  Hetio & FunCoup & \parbox{30pt}{\footnotesize\centering \hspace{6pt} StringDB \\(all)}  & Landmark \\ \midrule
	\multirow{7}{*}{\rotatebox{90}{\textbf{TCGA}}} & Coverage (\%)                 &  100 &      79 &   35 &   87 &   86 &   82 & 92 & 100 \\
	                                               & Sparsity (\%)                  &  0 &      \textbf{99.94} &   \textbf{99.94} &   \textbf{99.87} & \textbf{99.89} &    \textbf{98.66} & \textbf{98.61} & \textbf{95.18} \\
	                                               & Per-gene AUC SD             &   0.017  & 0.010  & 0.005  & 0.011  & 0.013  &  0.011 & 0.011  &  0.013  \\
	                                               & All Genes AUC ($\pm 2\mathrm{e}^{-4}$)  & 0.782 &      0.636 &   0.590 &   0.699 &   0.640 &    0.702 &           0.771 &     \textbf{0.788} \\

	                                               & All Genes improvement ($\pm 2\mathrm{e}^{-4}$) &   ---   &    -0.146 & -0.193 & -0.083 & -0.143 &  -0.081 &         -0.011 &    \textbf{0.006} \\ 
	                                               & Covered Genes AUC ($\pm 2\mathrm{e}^{-4}$) &   0.782  & 0.673  & 0.753  & 0.730  & 0.663 &  0.749 & \textbf{0.802}  & 0.788  \\ 
	                                               & Covered Genes improvement ($\pm 2\mathrm{e}^{-4}$)  &   ---  & -0.126  & -0.056  & -0.062  & -0.130 &  -0.045 & \textbf{0.013}  & \textbf{0.006}  \\ \midrule
	\multirow{7}{*}{\rotatebox{90}{\textbf{GTEx}}} & Coverage (\%)                 &  100 &    48 &  21 &   52 &   52 &   49 & 55 & 100 \\
	                                               & Sparsity (\%)                 &  0 &  \textbf{99.98} &   \textbf{99.98} &   \textbf{99.96} &  \textbf{99.96} & \textbf{99.53} &    \textbf{99.51} & \textbf{97.14} \\
	                                               & Per-gene AUC SD             &  0.036   &  0.012  & 0.005  &  0.012  & 0.015  & 0.015  & 0.013  &  0.029  \\
	                                               & All Genes AUC    ($\pm 2\mathrm{e}^{-4}$)      & 0.734 &      0.599 &   0.564 &   0.647 &   0.599 &    0.630 &           0.685 &     \textbf{0.748} \\

	                                               & All Genes improvement ($\pm 2\mathrm{e}^{-4}$) &    ---  &    -0.222 & -0.170 & -0.212 & -0.135 &  -0.105 &         -0.050 &    \textbf{0.014} \\ 
	                                               & Covered Genes AUC ($\pm 2\mathrm{e}^{-4}$)   &   0.734  & 0.707  & 0.802  & 0.781  & 0.693 & 0.764 & \textbf{0.837}  & 0.748  \\ 
	                                               & Covered Genes improvement ($\pm 2\mathrm{e}^{-4}$)  & ---    & -0.116  & -0.027  & -0.037  & -0.126 &  -0.054 & \textbf{0.021}  & \textbf{0.014}  \\  \bottomrule
\end{tabular}
}
\end{table*}


\subsection{Performance of the models}

\begin{figure}[!h]
    \centering
        \subfloat[TCGA]{\includegraphics[width=0.5\columnwidth]{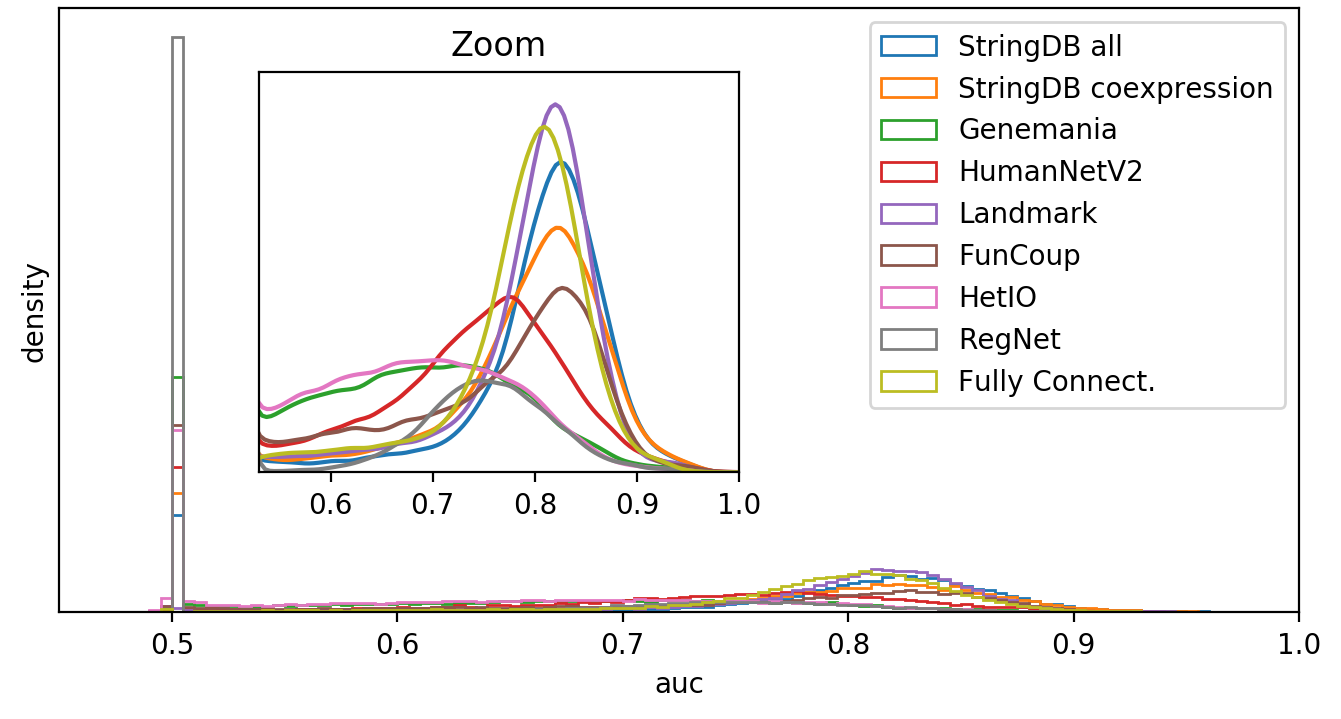}}
        \subfloat[GTEx]{\includegraphics[width=0.5\columnwidth]{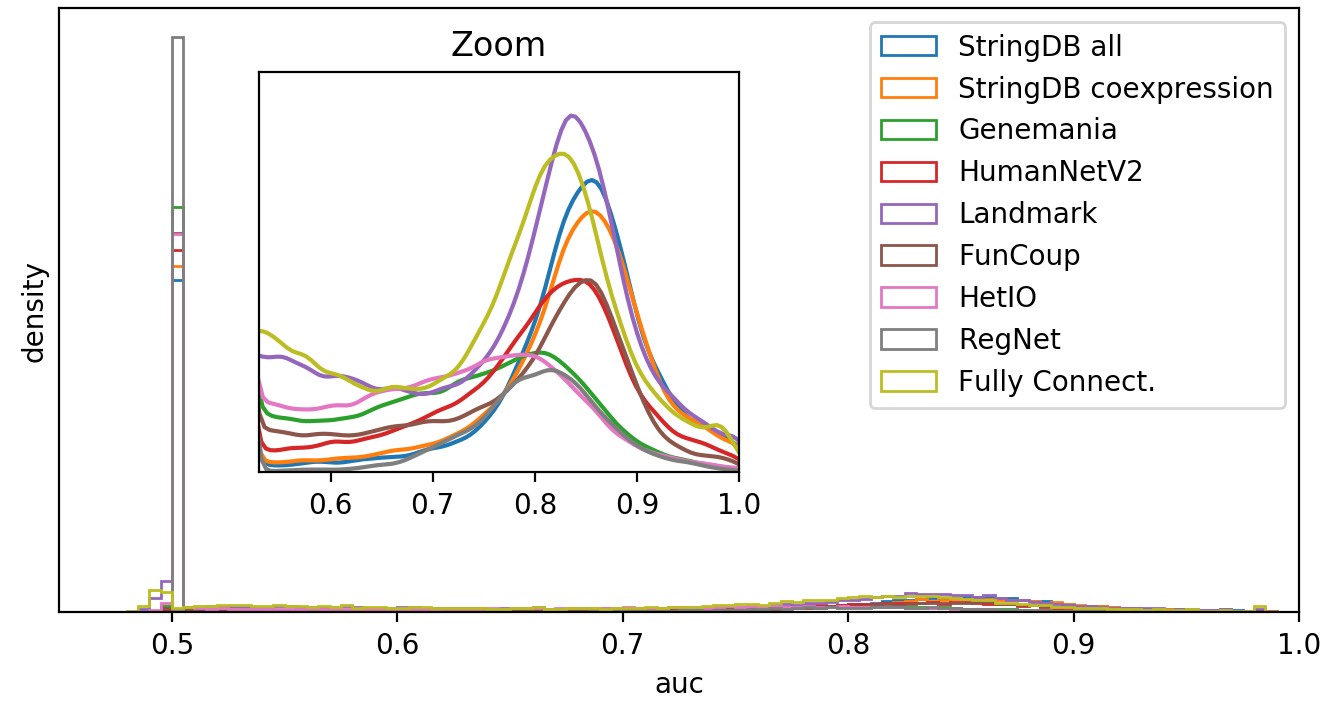}}
        \caption{Distribution of AUCs over genes, averaged over 3 trials, using different graphs. \textbf{Left}: TCGA dataset. \textbf{Right}: GTEx. The distribution is computed over all genes in the datasets.
        }
    \label{fig:auc_dist}
\end{figure}

We first visualize how well the different models fit, for the different graphs and genes. The distribution of the AUCs over the set of all genes, averaged across trials, for all the graphs is visualized in Figure \ref{fig:auc_dist}. For both datasets, there is a considerable spike at 0.5 which is mostly due to the fact that we assign a 0.5 AUC to the genes not present in the graph and some graphs have a very small proportion of genes in the datasets (especially for the GTEx dataset). However, some part of the peak at 0.5 AUC does correspond to covered genes for which the trained model actually achieved an AUC of 0.5, which could be due to either bad convergence or too restrictive feature selection. We also plot distributions of AUCs of the graphs over its set of covered genes (which differs between graphs) as supplementary material in Figure \ref{fig:auc_dist_intersec}.

We report the mean AUCs along with other statistics in Table \ref{tab:coverage} and also report the per-gene standard deviation of AUC across the three trials to assess the robustness of the evaluation.
It is noticeable that, overall, the best fit is not achieved using the fully connected graph. The best fit is obtained using the Landmark graph for both datasets if all the genes in the dataset are considered. Note that the Landmark graph has the highest coverage and lowest sparsity among all graphs but the fully connected one. If we only consider the performance on the set of covered genes, best performance are achieved using the StringDB. There is an anomaly in the case of RegNet which is consistently the worst graph on the set of all genes for both datasets due to its very low coverage. In general, predictive performance was better on TCGA than GTEx.

\subsection{Comparison with the fully connected graph}\label{comparefullyconnected}

\begin{figure}[!h]
    \centering
        \subfloat[TCGA]{\includegraphics[width=0.5\columnwidth]{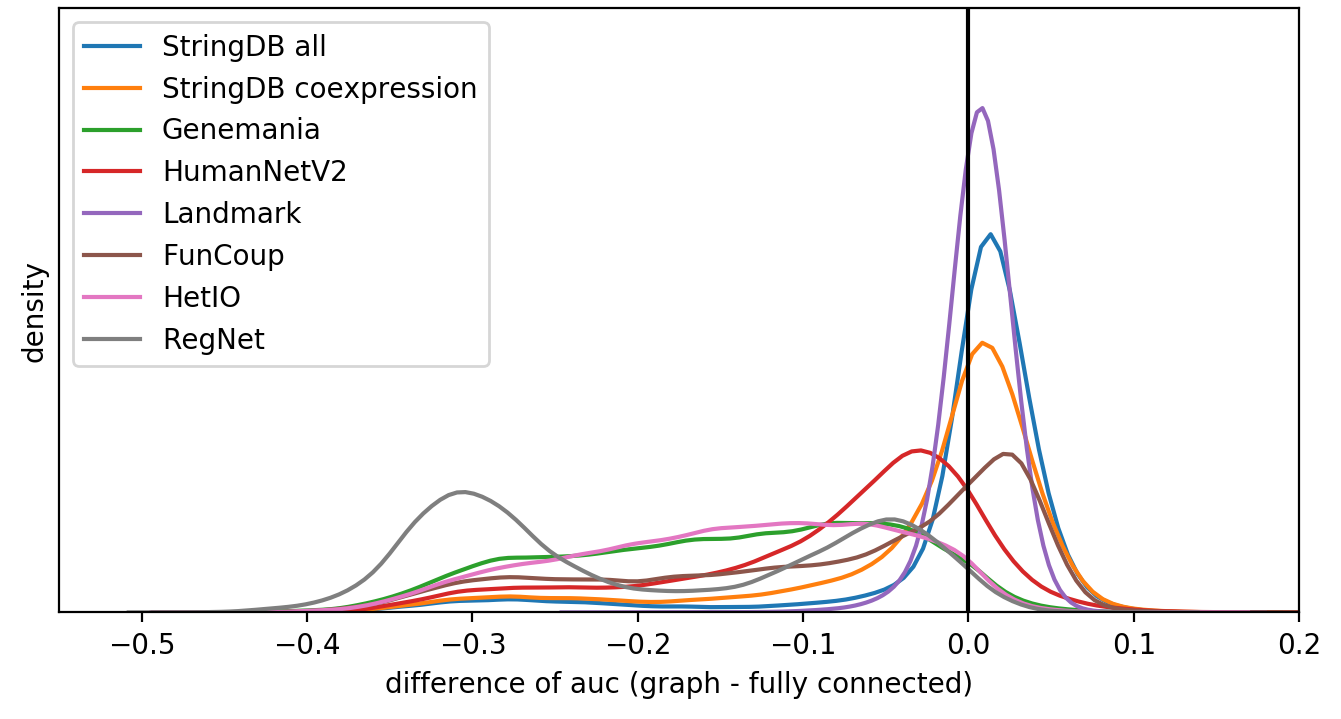}}
        \subfloat[GTEx]{\includegraphics[width=0.5\columnwidth]{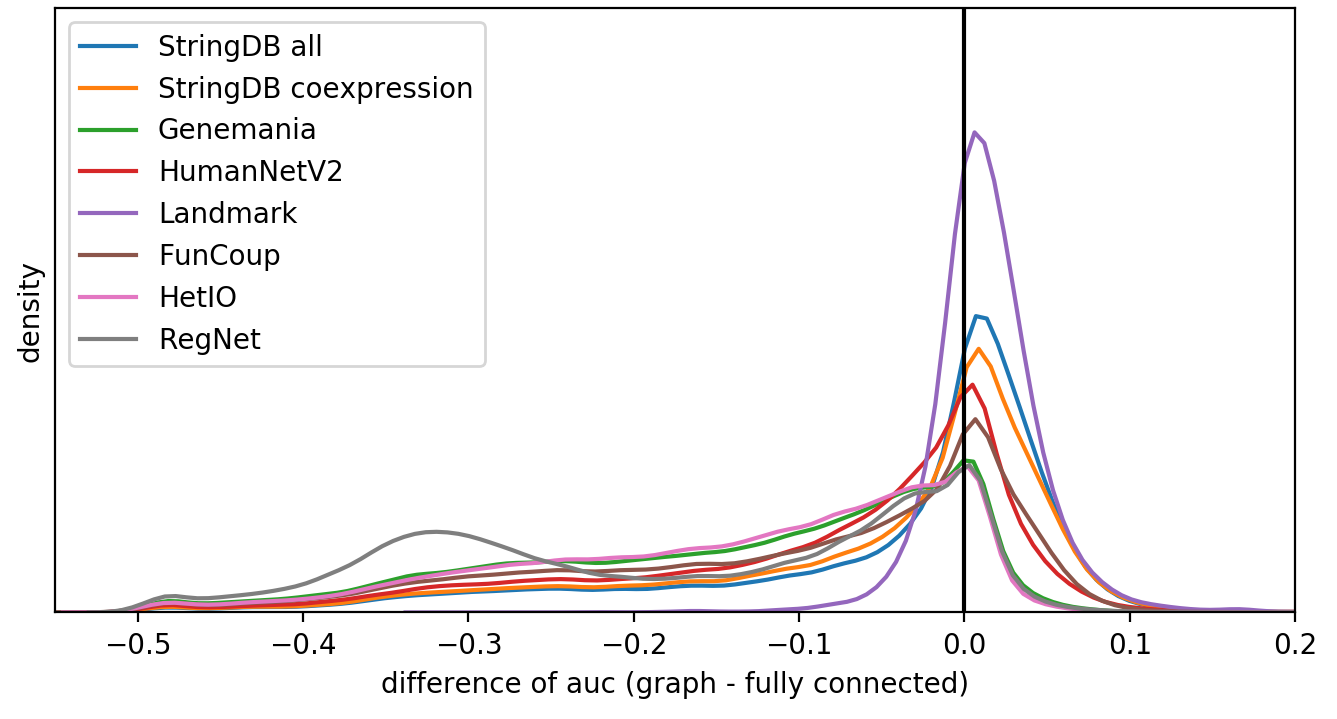}}
        \caption{Distribution of the All Genes AUC improvement relative to the fully-connected graph. \textbf{Left}: TCGA dataset. \textbf{Right}: GTEx. For each gene the difference in AUC is averaged over 3 trials. 
        }
    \label{fig:compare}
\end{figure}

For a given gene $g_i$, we define its \textit{AUC improvement} as  $AUC(neighbours_i) - AUC(all \; genes)$ which is the difference between the AUC of the model taking first neighbours as input and the AUC of the model taking all other genes as input. This difference  measures to which extent the inclusivity property holds, \textit{i.e.} the quality of the prior knowledge provided by the graph for gene $g_i$. The distribution of \textit{AUC improvements} over all genes is plotted in Figure \ref{fig:compare} where each difference has been averaged over three trials. Distributions of \textit{AUC improvements} over the set of Covered Genes are provided as supplementary material in Figure \ref{fig:compare_intersec}.

\begin{figure}[!h]
    \centering
    \includegraphics[width=0.7\columnwidth]{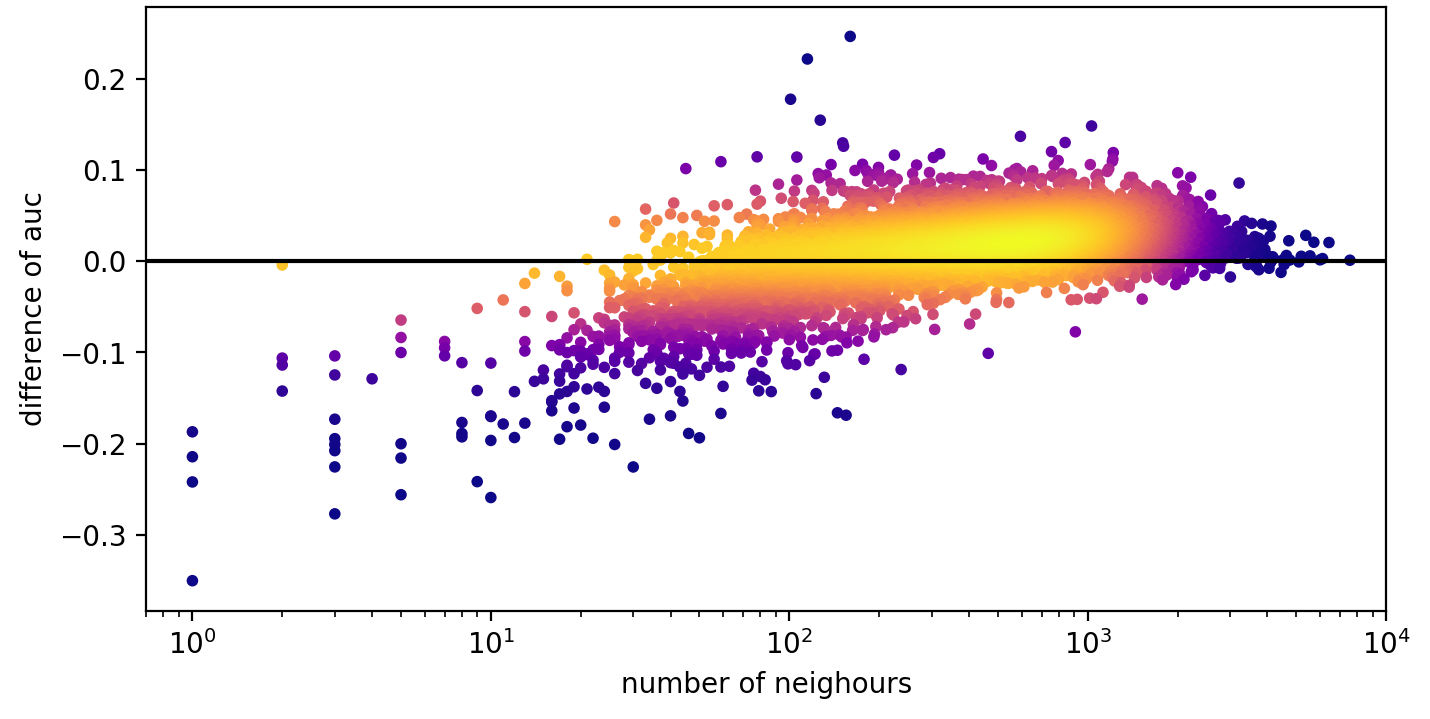}
    \caption{Covered Genes AUC improvement with respect to the number of neighbours, for StringDB (all) on the TCGA dataset. Color indicates density, with yellower being higher. Averaged over 3 trials.
    }
    \label{fig:diffauc_nneighb}
\end{figure}

For graphs such as StringDB and Landmark, a most of the mass of the distribution is close to zero or in the positive region, meaning that the equality (Eq. \ref{eq:1}) holds for a most genes. For those same graphs, the distribution has an heavy tail in the negative area, corresponding to genes with highly negative \textit{AUC improvement} for which the equality (Eq. \ref{eq:1}) is not satisfied. Note that StringDB and Landmark have high coverage and lower sparsity than most other graphs. Those failure cases should be the consequence of missing edges in the graphs. To confirm this hypothesis, we plot the distribution of per-gene AUC differences with respect to the number of neighbours of genes in Figure \ref{fig:diffauc_nneighb}. Highly negative AUC differences seem to correspond to genes with a relatively small number of neighbours. Figure \ref{fig:diffauc_nneighb} also highlights that when the number of neighbours is sufficient, most genes have a positive \textit{AUC improvement}. Removing redundancy of features, while keeping relevant information allowed models to achieve slightly better performance. For other graphs such as HetIO and GeneMANIA, the high majority of the mass is in the negative region, meaning that the equality (Eq. \ref{eq:1}) does not hold for most genes.

\subsection{Random R-\textit{n} graphs}

We would like to assess how much of the graph's internal structure contributes to its performance in our evaluation by comparing it to random graphs. We generate graphs with a fixed number of randomly sampled neighbours \textit{n} from the set of genes in the dataset and evaluating them on the task. For each target gene in the dataset, we would sample \textit{n} other genes from the dataset and connect them to the target gene node to create an R-\textit{n} graph. We created 15 such graphs with \textit{n} varying between 10 and 10,000 and ran three trials with randomly sampled neighbours for every trial instead of keeping them fixed for all three trials. 

We plot the improvement in AUC for some of these graphs in Figure \ref{fig:rand-diff-auc} of the supplementary material. As expected, when the number of random neighbours \textit{n} increased, the feature selection related to the graphs achieved better performance. R-\textit{n} graphs perform on par or better than the baseline on average for \textit{n} greater than $500$, meaning that using 500 randomly sampled features actually performs on par or better with using the entire gene set. The standard deviation across trials of the per-gene AUC is $1\mathrm{e}^{-2}$. There is a bump of well predicted genes even when \textit{n} is small, for the GTEx dataset only. One explanation could be that GTEx is mostly composed of healthy cells which makes the population of cells more homogeneous and predictions easier for some genes, but further investigation is needed to confirm this hypothesis.

\subsection{A map of graphs}

\begin{figure}[!htbp]
    \centering
        \subfloat[TCGA]{\includegraphics[width=0.6\columnwidth]{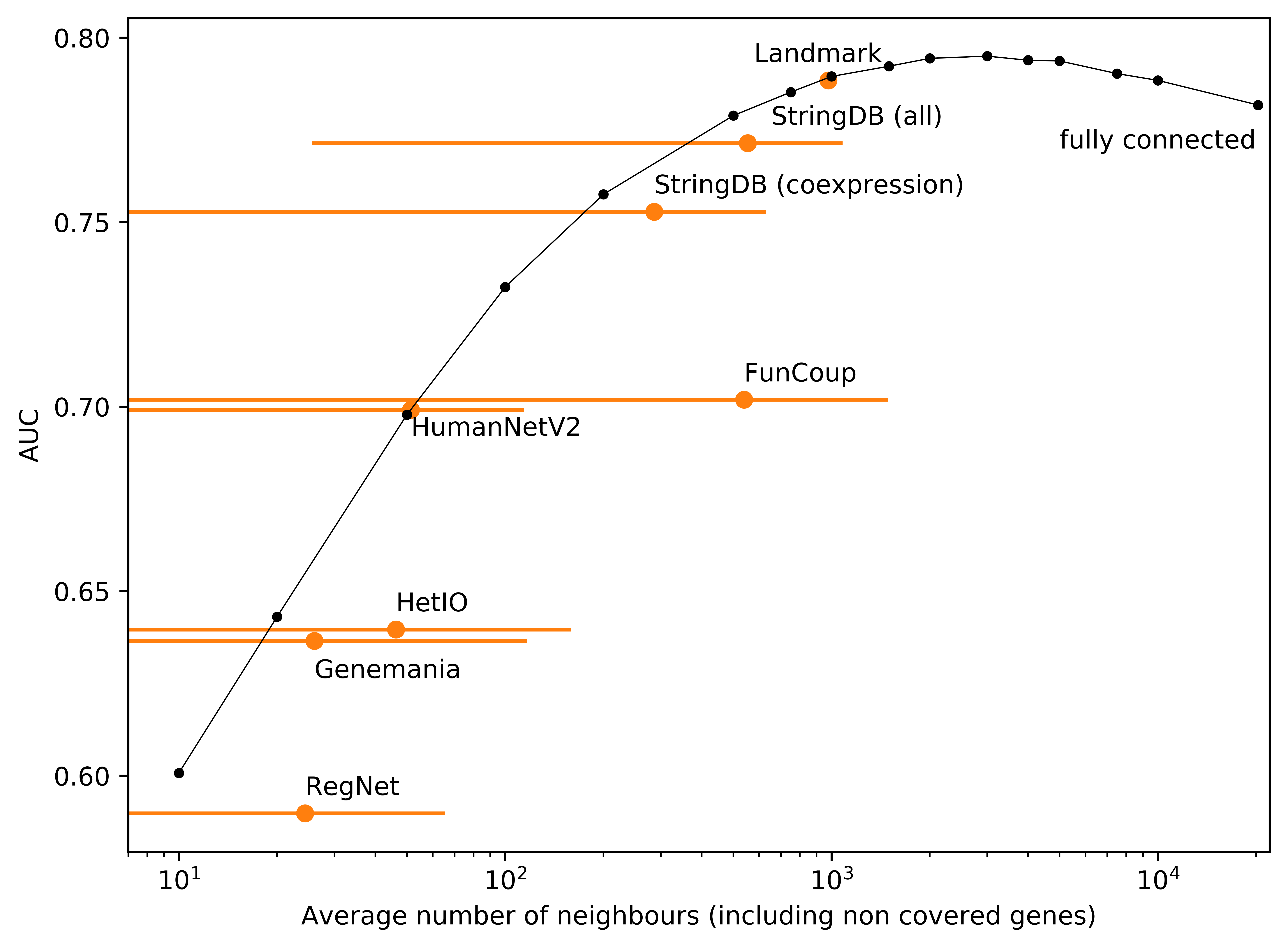}}
        \qquad
        \subfloat[GTEx]{\includegraphics[width=0.6\columnwidth]{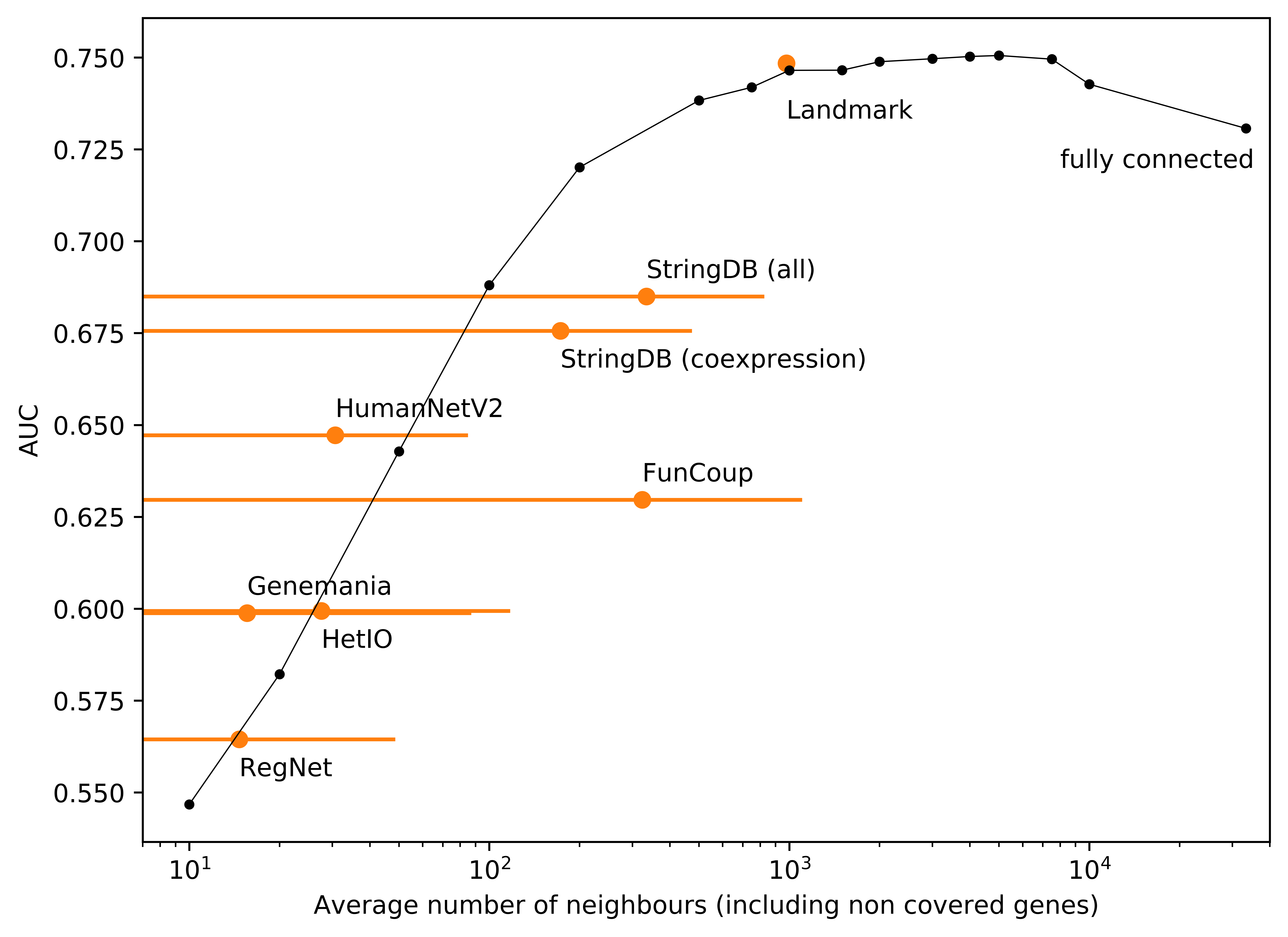}}
        \caption{Mean AUC on all genes as a function of average number of neighbours in the graph (logarithmic scale). The black line represents the average AUCs achieved with the multiple R-\textit{n} graphs and one could think about it as the level of randomness. The orange lines indicate the standard deviation of the number of neighbours for each graph. The standard deviation of AUCs is $2\mathrm{e}^{-4}$ and too small to be plotted.}
    \label{fig:aucVneighborsall}
\end{figure}

Figure \ref{fig:aucVneighborsall} shows the mean AUC over genes as a function of the average number of neighbours in the graph. The mean AUC is computed over all genes in the dataset. Results on the set of genes covered by all graphs are presented as supplementary material in Figure \ref{fig:aucVneighbors}. The R-\textit{n} graphs are depicted as a black line which could be thought of as the level of randomness. On the set of all genes, almost all the curated graphs are below the level of randomness except for HumanNetV2 (GTEx), GeneMANIA (GTEx) and Landmark. Poor performance is in part due to the limited coverage of some of the graphs, in which some genes do not have any neighbours that made the prediction a random guess. On the set of genes covered by all graphs, most curated graphs are above the level of randomness, but only StringDB achieves better performance than the best performing random graph, and even then only by a very small margin ($\leq 5 \mathrm{e}^{-3}$ AUC).

\subsection{Experiments with clinical tasks}

In order to support our evaluation, we evaluated on different clinical tasks a one hidden layer, sparsely connected neural network whose connections are based on the prior knowledge provided by the different curated graphs. The number of hidden units was chosen to be equal to the dimension of the input, and ReLU activation functions were used. The weights of the first layer are masked with $\Tilde{A} = A + Id$ where $A$ is the adjacency matrix of the graph and $Id$ the identity matrix. It means that the activation of the first hidden unit is a feature computed from the first gene and its neighbours only. Even if this model does not correspond to state of the art graph neural networks, it allowed us to make a simple comparison between sparsified models and an unbiased (fully connected) one. We trained each model with 5-folds cross-validation for 100 epochs on the set of genes covered by all graphs (but RegNet). We used the Adam optimizer with a learning rate equal to $1\mathrm{e}^{-6}$. We only optimized the learning rate for the fully connected model on the ('PAM50Call\_RNAseq', 'BRCA') task. 

\begin{figure}
    \centering
    \includegraphics[width=0.7\columnwidth]{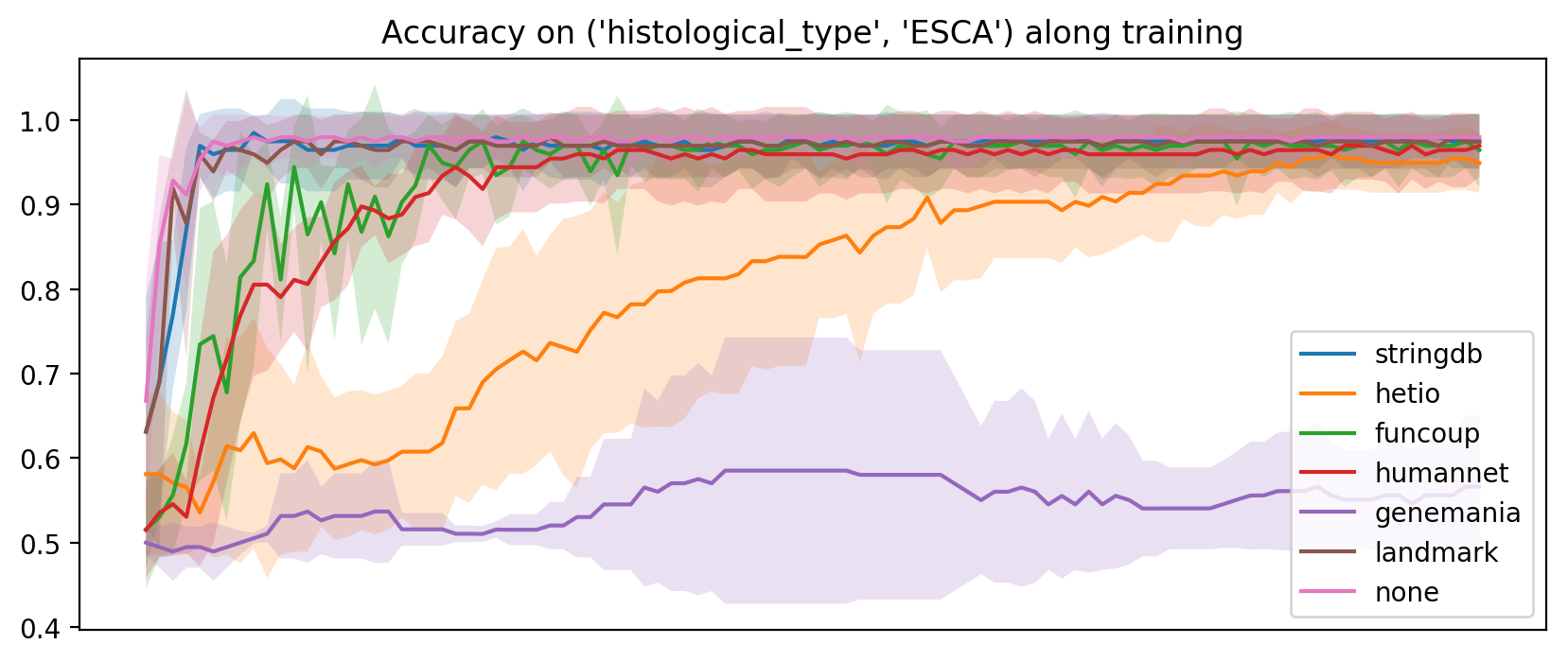}
    \caption{Test accuracy on the set of genes covered by all graphs (\textit{Intersection Set}) along training. The task is to predict the histological type of Esophageal Carcinoma. The reported accuracy is an average over 5 folds of cross-validation. Shaded areas represent the standard deviation across folds.}
    \label{fig:clinicaltask_histo}
\end{figure}

We evaluated on five tasks taken from the \href{https://github.com/mandanasmi/TCGA_Benchmark}{TCGA Benchmark} \citep{m2019tcga}. These particular tasks were chosen because they have shown to allow improvement compared to the majority class predictor. Each task was cast as a binary classification (majority class versus all). Test accuracies along training are reported in Figure \ref{fig:clinicaltask_histo} for the prediction of histological type in Esophageal Carcinoma, and as supplementary material in Figure \ref{fig:clinicaltasks} for the other tasks. 

Overall, the comparative performance between graphs are consistent with our map of graphs shown in Figure \ref{fig:aucVneighborsall}. The model without prior knowledge (fully connected) consistently outperforms the other models. The relative performance between graphs, both in terms of speed of adaptation and final accuracy, seem to be consistent with our analysis. Second best performance is achieved using the STRINGdb and Landmark graphs. FunCoup and HumanNet follow, before Hetionet and finally GeneMANIA.  The poor performance of GeneMANIA could be related to its extreme sparsity. Note that for some tasks, some models do not converge well, this is a well known phenomenon observed when training sparsely connected neural networks \citep{frankle2018lottery, Cohen2016}.

\section{Discussion}

We proposed a property that gene interaction graphs should satisfy in order to provide ``good'' prior knowledge as we defined it, in the context of predictions on gene expression data. A Single Gene Inference task was used to test whether the property holds for a given graph. This test provides a measure for each gene and graph, that reveals whether the graph captures the right dependencies for this gene. 

We studied several existing curated graphs and found that large graphs such as StringDB and Landmark allow the equality (Eq. \ref{eq:1}) to hold for most genes, while being very sparse. This suggests that using the entire gene set to predict the expression of a gene tends to include mostly uninformative features as the expression of most genes can be explained with a fraction of the gene set. Those results validate the findings of \citet{Subramanian2017_s}, that a lot of information is shared across gene expressions


We then compared existing curated graphs against randomly generated graphs to assess the usefulness of the prior knowledge they provide. We found that randomly selecting 500 or more genes as neighbours for a target gene can perform on par with or improve over the baseline for most genes in TCGA and GTEx. This means that a random graph R-\textit{500} allows the equality (Eq. \ref{eq:1}) to hold for most genes, while being quite sparse. Those results confirm that the relevant information about the state of the cell is spread across many genes. Future work could explore random graphs that preserve the statistics (\textit{e.g.} distribution of degrees) of a given graph, in order to have a more precise comparison between curated graphs and their random counterparts. 

We finally validate those results by comparing the performance of models inputted with prior knowledge coming from the different curated graphs. Relative performances are consistent with our analysis of the goodness of the prior knowledge provided by the different graphs.

There are several limitations to this pipeline. We test whether the graph captures the right dependencies in the data, which is not restricted to causal dependencies. For example, consider three genes $A$, $B$ and $C$ where $A$ downregulates $B$ and $B$ downregulates $C$ ($A \rightarrow B \rightarrow C$) and suppose no noise or randomness is added in the process ie the links between these genes are deterministic. Our pipeline will not be able to distinguish between the true causal graph and any other graph where $A$, $B$ and $C$ are connected together (in fact it is not possible to do so from observational data). Maybe the curated graphs contain causal information that cannot be recovered from the raw observations. 

Moreover, we did not take into account the noise in the observations of gene activations, which is due to limitations in the acquisition process of gene expression data. Suppose one is provided with the \textit{true} graph and that our data corresponds to noisy observations of the \textit{true} activation of genes. Second and higher-degree neighbours can still have a denoising effect and give information on the \textit{true} level of expression of a gene. Let us consider the results from \citet{Dutil2018}. For genes they report, taking into account second and higher degree neighbours did not yield any improvement compared to only first degree neighbours. Nonetheless, further investigation would be needed in order to assert that one does not need to take into account the noise of the data to test the goodness of graph-based prior knowledge.

To summarize, the additive value of using curated graphs to provide prior knowledge appears to be limited. Nonetheless, curated graphs could be valuable when one is interested in specific subgroups of genes which have been well studied by biologists. This analysis is left for future work.

%% file: sections/appendix.tex
\clearpage
\appendix
\onecolumn

\newlength{\appendixfigwidth}
\setlength{\appendixfigwidth}{0.7\columnwidth}

\renewcommand\thesection{\Alph{section}}

\renewcommand\thefigure{\thesection.\arabic{figure}}
\renewcommand{\thepage}{\thesection.\arabic{page}}

\setcounter{page}{0}
\setcounter{figure}{0}

\section{Appendix: Supplementary figures}

\begin{figure}[h!]
    \centering
        \subfloat[TCGA]{\includegraphics[width=\the\appendixfigwidth]{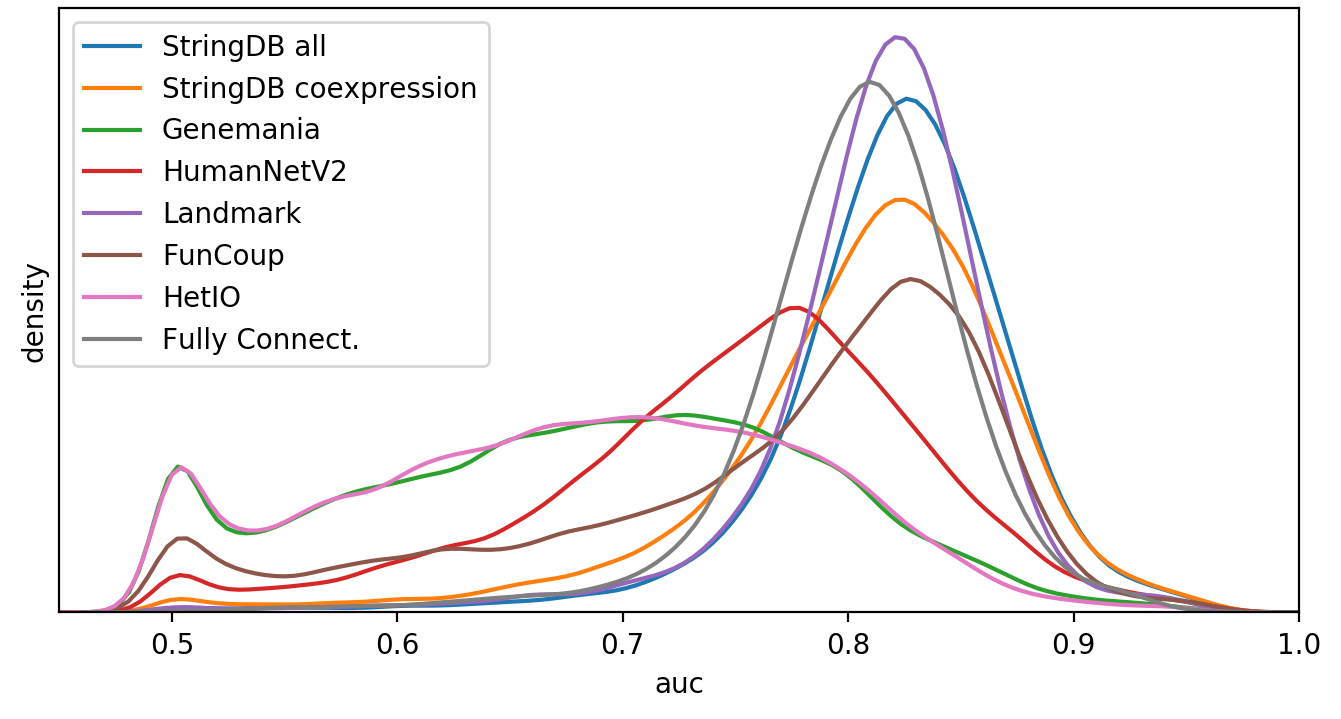}}
        \qquad
        \subfloat[GTEx]{\includegraphics[width=\the\appendixfigwidth]{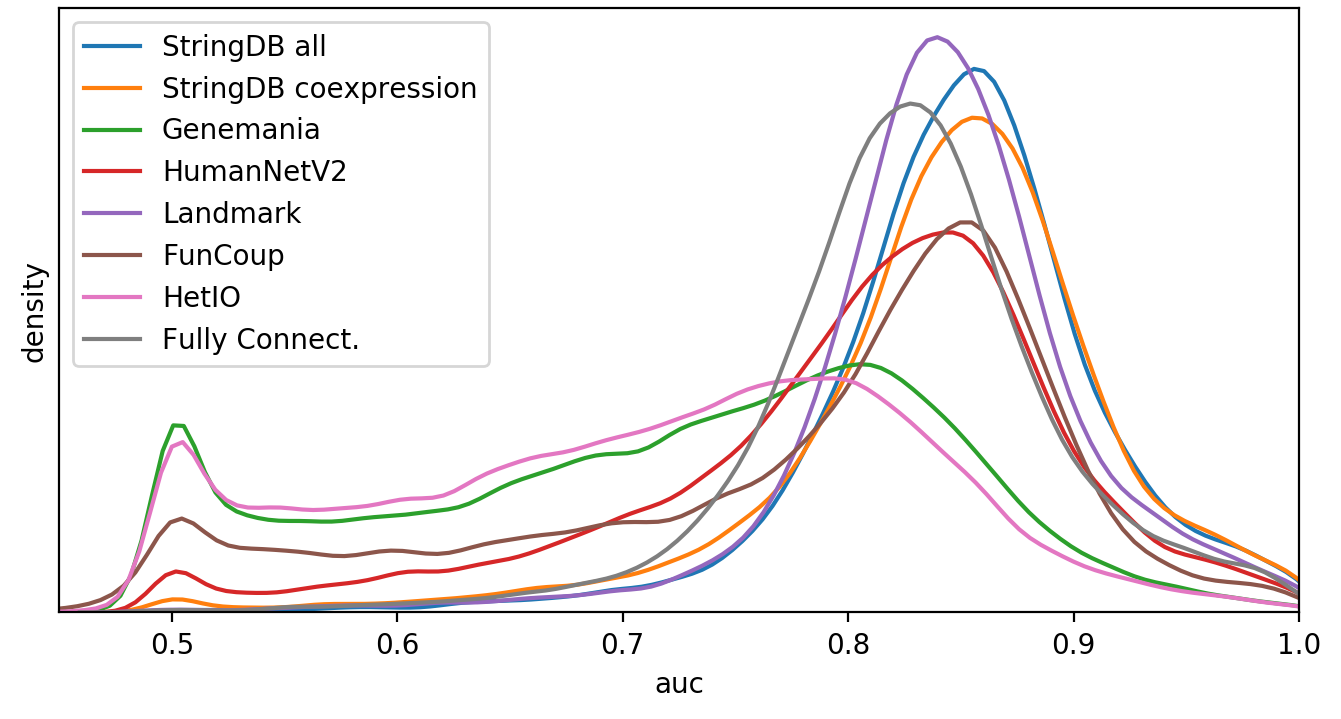}}
        \caption{Distribution of Covered AUCs, averaged over 3 trials, using different graphs. \textbf{Top}: TCGA dataset. \textbf{Bottom}: GTEx. The distribution is computed over genes covered by all graphs except RegNet (14442 genes for TCGA and 14270 for GTEX).
        }
    \label{fig:auc_dist_intersec}
\end{figure}

\begin{figure}[h]
    \centering
        \subfloat[TCGA]{\includegraphics[width=\the\appendixfigwidth]{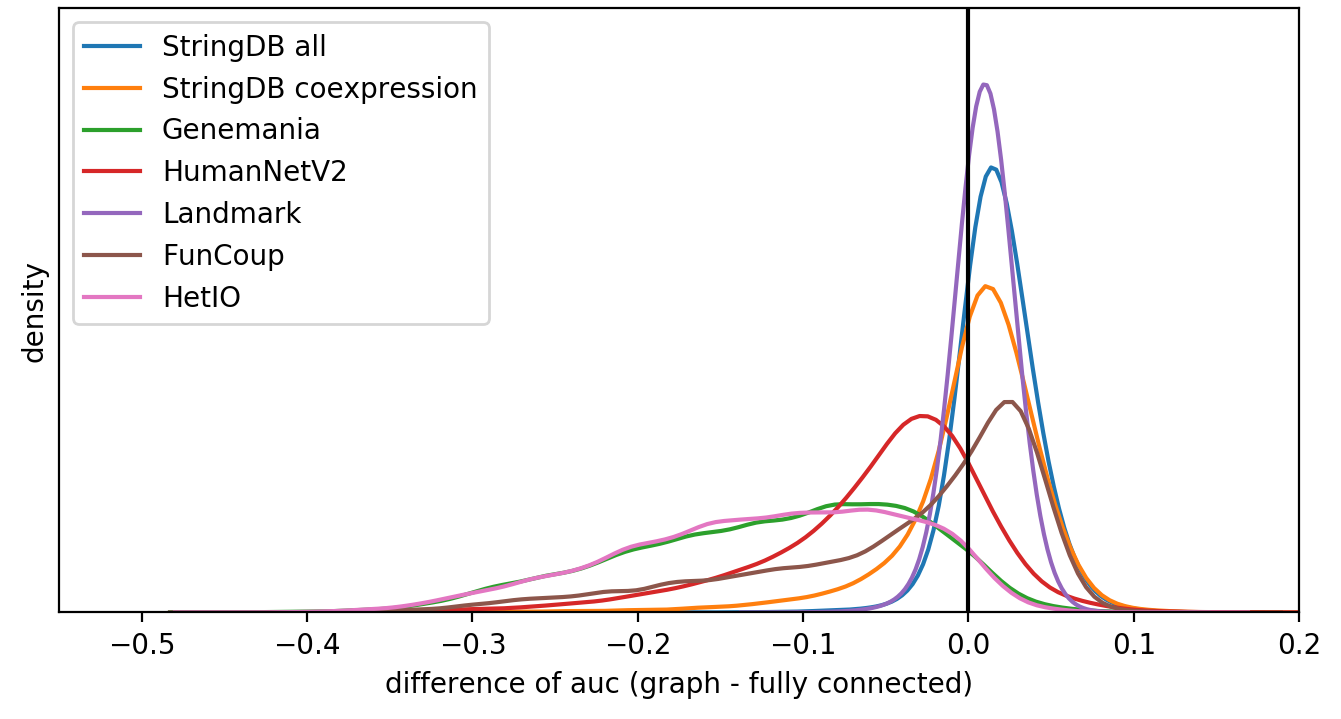}}
        \qquad
        \subfloat[GTEx]{\includegraphics[width=\the\appendixfigwidth]{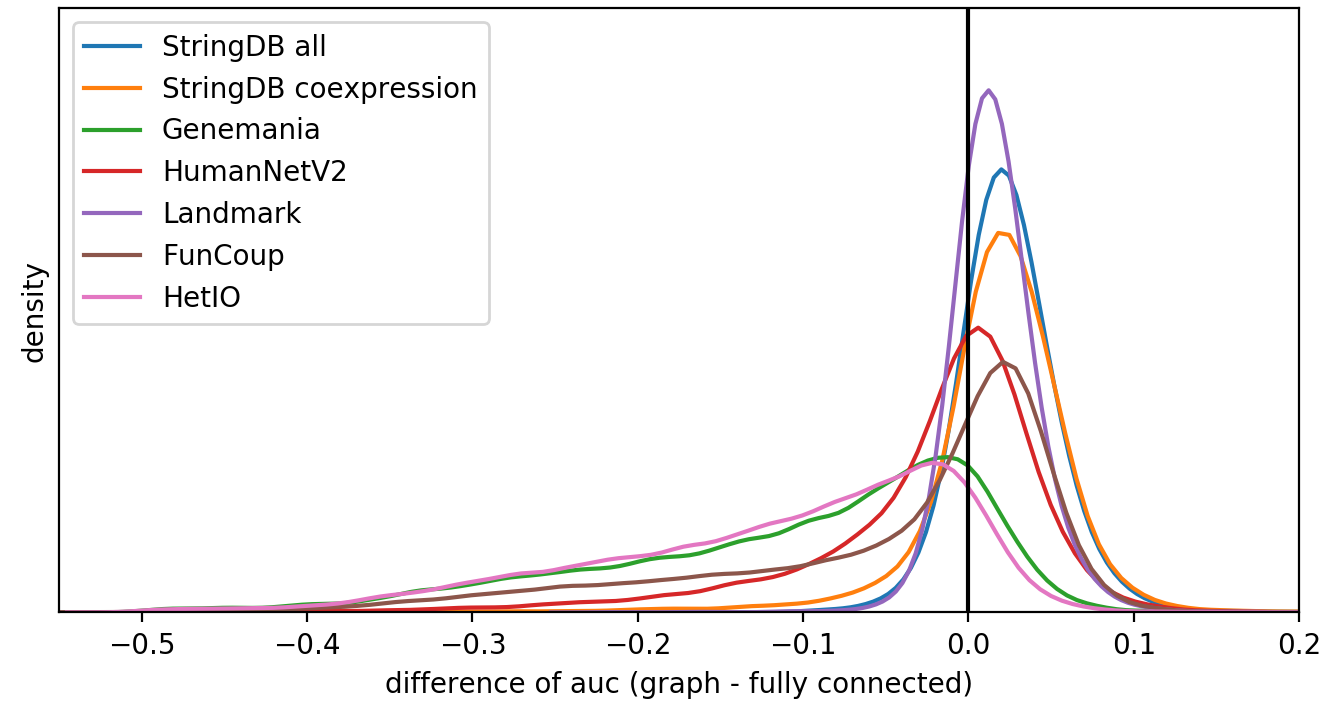}}
        \caption{Distribution of the improvement relative to the fully-connected graph on the set of genes covered by all graphs. \textbf{Top}: TCGA dataset. Intersection set composed of 14442 genes \textbf{Bottom}: GTEx. Intersection set composed of 14270 genes. For each gene the difference in AUC is averaged over 3 trials. 
        }
    \label{fig:compare_intersec}
\end{figure}

\begin{figure}[h]
    \centering
        \subfloat[TCGA]{\includegraphics[width=\the\appendixfigwidth]{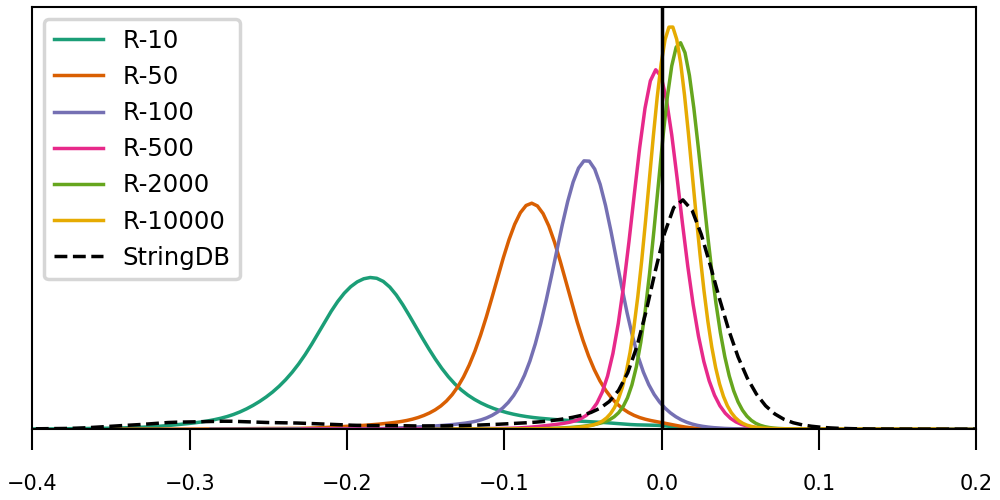}}%
        \qquad
        \subfloat[GTEx]{\includegraphics[width=\the\appendixfigwidth]{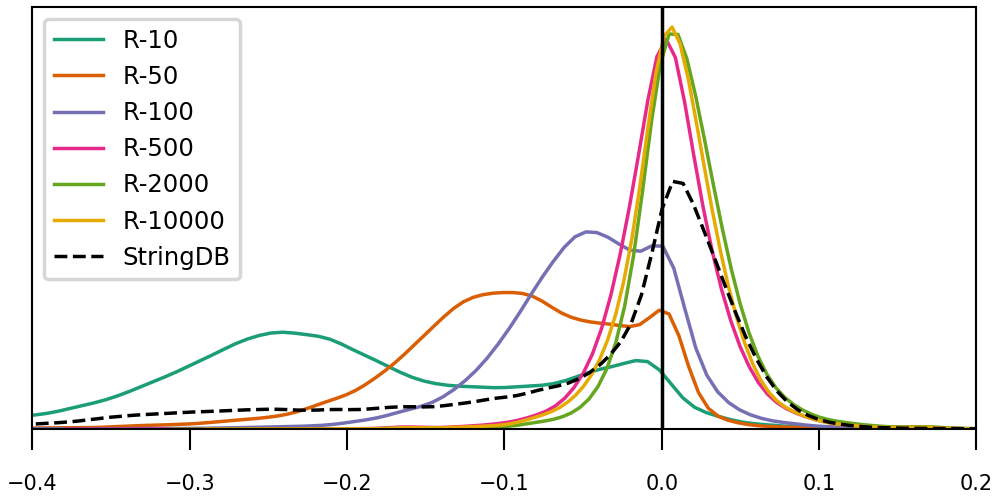}}    
        \caption{Distribution of AUC improvements over all genes in R-\textit{n} graphs. The dashed line represents StringDB, the best performing curated graph. The more towards the right the better the performance.
    }
    \label{fig:rand-diff-auc}
\end{figure}

\begin{figure}[h]
    \centering
        \subfloat[TCGA]{\includegraphics[width=\the\appendixfigwidth]{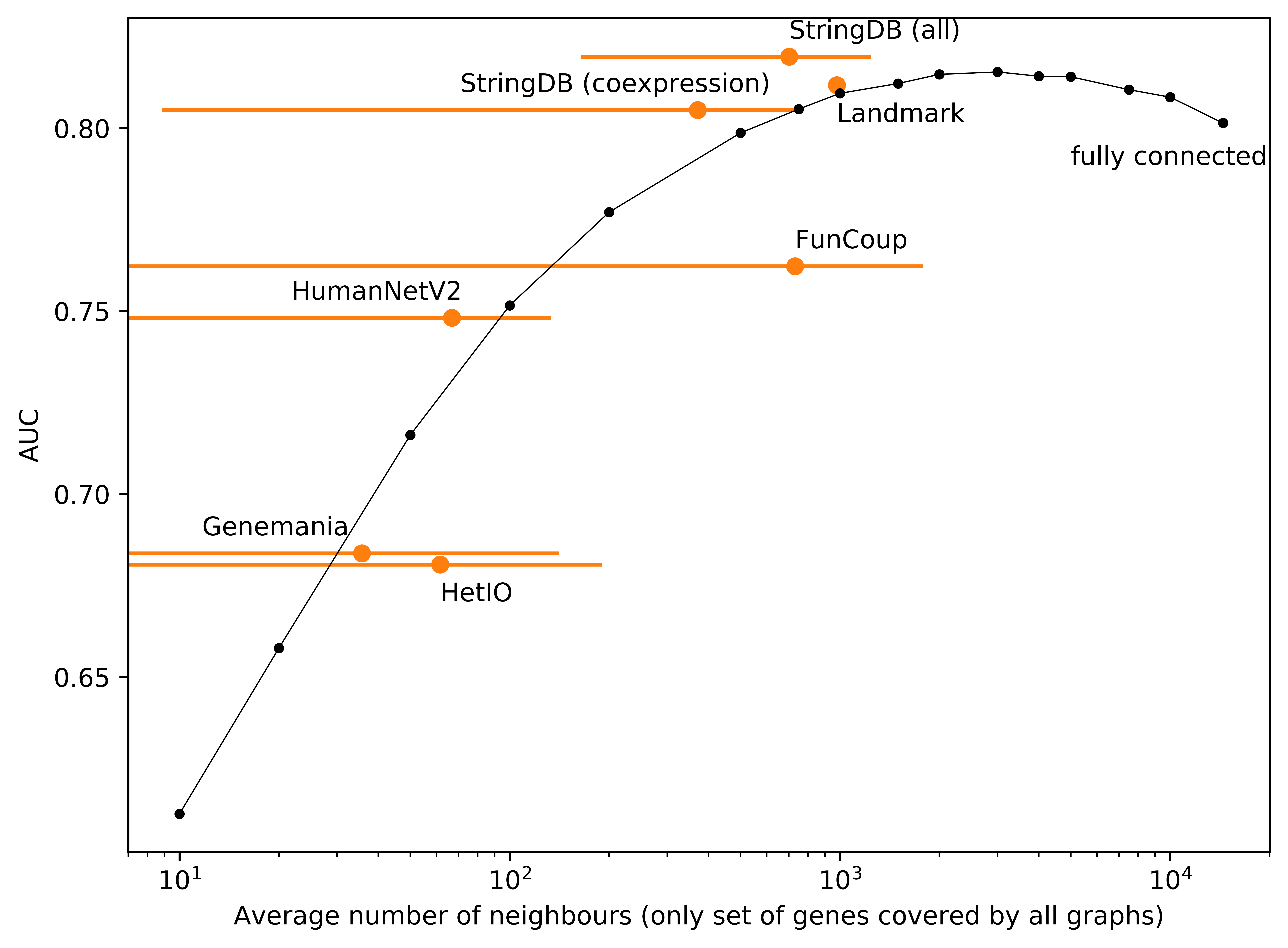}}
        \qquad
        \subfloat[GTEx]{\includegraphics[width=\the\appendixfigwidth]{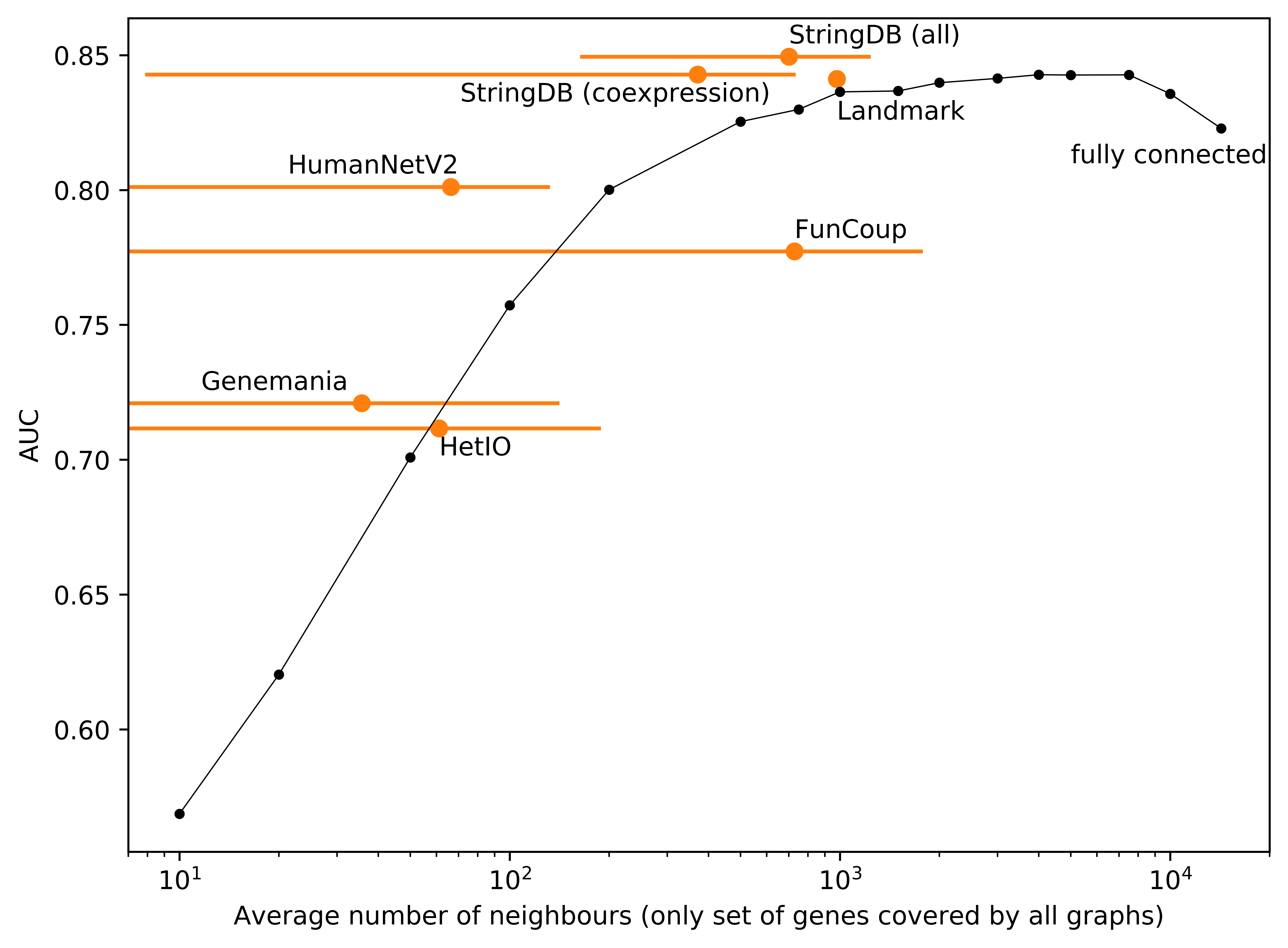}}
        \caption{Mean AUC on the set of genes covered by all graphs (\textit{intersection set}) as a function of average number of neighbours (on the \textit{intersection set}) on a logarithmic scale. The line represents the average AUCs achieved with the multiple R-\textit{n} graphs, and one could think about it as the level of randomness. Horizontal bars represent the standard deviation of the number of neighbours. For TCGA, there are 14445 genes in the intersection set, and 14270 for GTEx.}
    \label{fig:aucVneighbors}
\end{figure}

\begin{figure}[h]
    \centering
    \includegraphics[height=0.7\textheight]{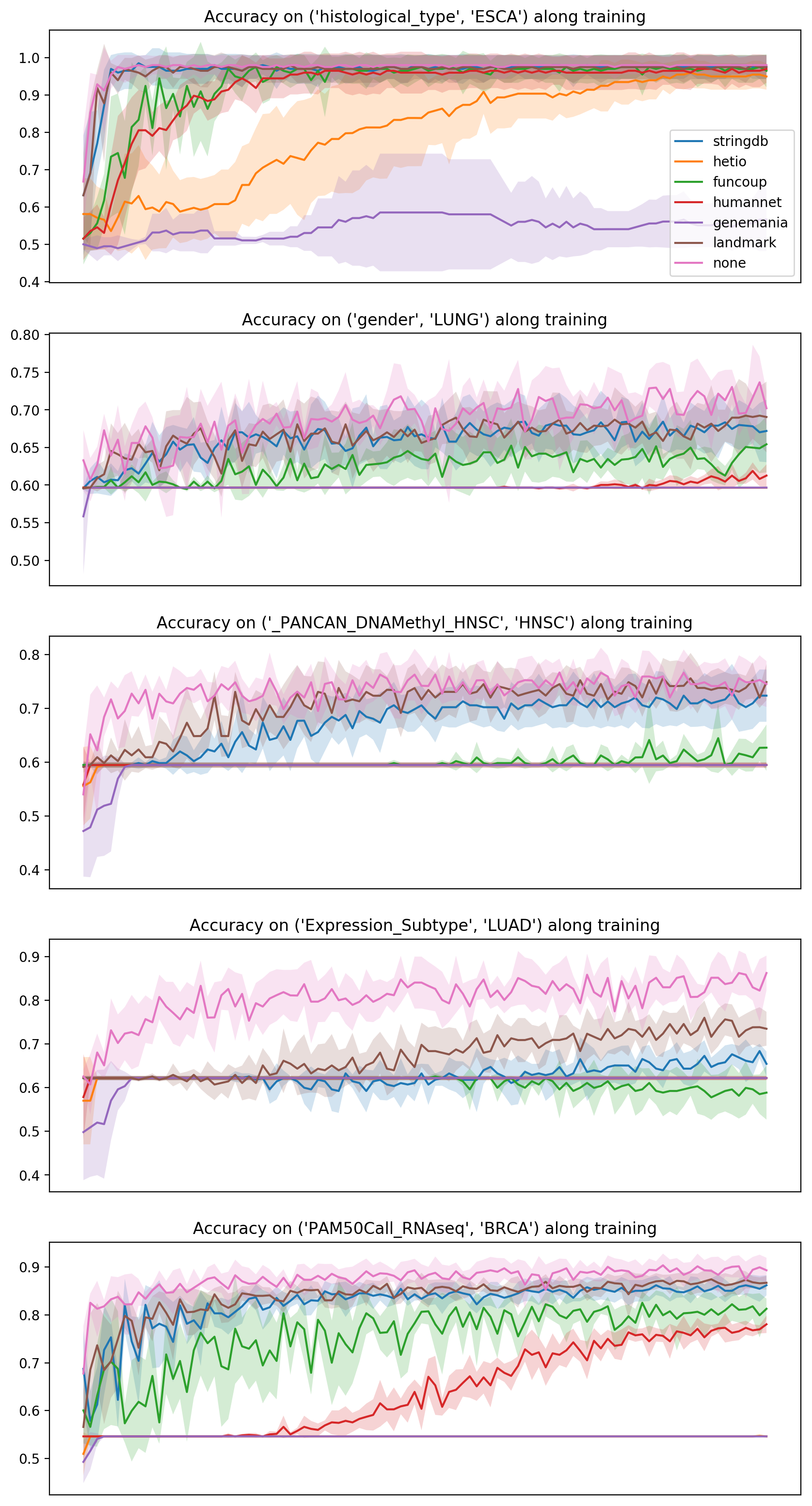}
    \caption{Test accuracy on the set of genes covered by all graphs (\textit{intersection set}) along training for 5 different tasks. The reported accuracy is an average over 5 folds of cross-validation. Shaded areas represent the standard deviation across folds. The model without prior knowledge consistently outperforms the others. Overall, the relative performance between graphs, both in terms of speed of adaptation and final accuracy, seem to be consistent with our analysis. The very bad performance of \textit{GeneMANIA} could be related to its extreme sparsity.}
    \label{fig:clinicaltasks}
\end{figure}